\documentclass{sig-alternate}
\pdfoutput=1

\usepackage{amsmath, graphicx}
\usepackage[ruled,vlined]{algorithm2e}
\usepackage{mathtools}

\newcommand{\argmin}{\operatornamewithlimits{argmin}}

\begin{document}

\title{POSTER: Privacy-preserving Indoor Localization}

\numberofauthors{1} 

\author{
\alignauthor
Jan Henrik Ziegeldorf, Nicolai Viol, Martin Henze, Klaus Wehrle\\
       \affaddr{Communication and Distributed Systems (COMSYS), RWTH Aachen University, Germany}\\
       \email{\{ziegeldorf, viol, henze, wehrle\}@comsys.rwth-aachen.de}
}
\maketitle

\begin{abstract}
Upcoming WiFi-based localization systems for indoor environments face a conflict of privacy interests:
Server-side localization violates location privacy of the users, while localization on the user's device forces the localization provider to disclose the details of the system, e.g., sophisticated classification models.
We show how Secure Two-Party Computation can be used to reconcile privacy interests in a state-of-the-art localization system.
Our approach provides strong privacy guarantees for all involved parties, while achieving room-level localization accuracy at reasonable overheads.
\end{abstract}

\section{Introduction}

Nowadays, WiFi-based localization systems are increasingly considered for indoor environments.
Users only need a WiFi-capable device to repeatedly supply received signal strength indicator (RSSI) measurements, which are matched against a signal propagation model on the server side.
Thus, these approaches promise better usability than approaches that require users to carry special devices.
However, the system provider can track the user which gives rise for serious concerns, especially when deployed in sensitive areas such as hospitals.
To overcome these concerns, the localization could be performed directly on the user's device (similar to GPS).
As indoor environments are more challenging than outdoors, e.g., due to heavy multi-path signal propagation, sophisticated signal propagation models and carefully calibrated localization algorithms are required to achieve reasonable localization accuracy.
However, this information is potentially private to the system provider, e.g., the business secret of a small start-up developing indoor localization systems.
The user's concern for location privacy and the business interests of the system provider leave us with a dilemma and the problem statement of this paper:
How can the system provider localize the user i) without learning the user's location (\emph{user privacy}) and ii) without the user learning the provider's localization models (\emph{server privacy})?

In this paper, we show how to realize a state-of-the-art localization system based on Hidden Markov Models (HMMs) \cite{viol2012hidden} in a privacy-preserving way using Secure Two-Party Computation (STPC).
As a partial result, we outline how to implement the Viterbi decoder for HMMs more efficiently than previous works~\cite{aliasgari2013secure}. 
For the first time, we also present a prototype implementation and preliminary evaluation results which show that our approach achieves room-level accuracy at a localization frequency of one per 10 seconds, which is practical for many indoor location-based services.

\section{System \& Protocol Design}

The system presented in \cite{viol2012hidden} models a user's unknown sequence of positions as states of a HMM and uses the Viterbi decoder to determine the most likely path based on the user's supplied RSSI measurements.
The authors show that the algorithm is more accurate than simple nearest-neighbour approaches, for which privacy-preserving variants have been widely discussed \cite{rane2013privacy}.
Thus, we consider the challenge to redesign the HMM-based approach from \cite{viol2012hidden} for privacy.

\begin{figure}
\includegraphics[width=\columnwidth]{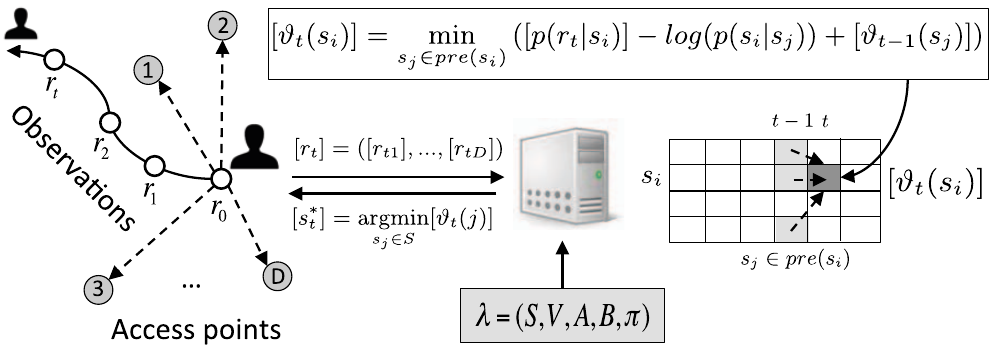}
\caption{\small{Overview of the localization system. Values in brackets are encrypted and computed using STPC.}}
\label{fig:system}
\vspace{-1em}
\end{figure}

Figure \ref{fig:system} gives a short overview of our localization system and protocol.
As the user moves around the scene, she periodically makes observations, i.e., RSSI measurements $r_{t,t=1..T}$ from $D$ access points (APs) in her vicinity.
She encrypts the observations and sends them to the server.
The server has trained a HMM $\lambda$ that can be used to iteratively infer the user's most likely position $s^*_t \in S$ based on her observations $r_{t,t=1..T}$.
This optimization problem is solved using the well known Viterbi algorithm:
In each timestep $t = 0...T$, we compute the probabilities $\vartheta_t(s_i)$ of being in state $s_i$ after having observed $r_0,...,r_t$.
Due to the Markov property, $\vartheta_t(s_i)$ depends only on the previous state and can be derived as shown in Figure \ref{fig:system}.
After each step, we send the most probable current position $s^*_t$ to the user.
Note that for technical reasons, we work in negative log-space thereby transforming the maximization to minimization problem, where the $\vartheta_t(s_i)$ values represent costs.

The challenge now is to carry out the described computations on the user's \emph{encrypted} inputs.
Here, we make use of STPC protocols.
STPC allows two parties with private inputs $x$ and $y$ to compute a functionality $f(x,y)$ without revealing anything to each other except the result of the computation.
In our case, the desired functionality is the Viterbi algorithm and the output a sequence of encrypted positions that only the user can decrypt.
To implement this, we build on standard STPC protocols based on homomorphic encryption \cite{lagendijk2013encrypted, rane2013privacy}.
The homomorphic property allows to perform additions and scalar multiplications on cipher texts efficiently without interaction.
Other operations, such as multiplication of cipher texts as well as minimum selection, require interaction with the user's device and are computationally more expensive.
This needs to be considered in an efficient protocol design.
We now describe in more detail how to execute the algorithm's steps privately using STPC.

\textbf{Inputs.} 
The private inputs of the user are the periodic RSSI measurements $r_t = (r_{t1}, ..., r_{tD})$ taken from $D$ APs.
The server's private input is the HMM $\lambda = (S, V, A, B, \pi)$ modelling the indoor environment.
The user encrypts her measurements before sending them to the server to protect her privacy.
The server uses its own inputs in clear but blinds any intermediate values during interactive protocol steps to protect the parameters of the HMM.
We write encrypted or blinded values in square brackets.

\textbf{Emission probabilities.} 
Probabilities $p(r_t|s_i)$ are modelled as multi-variate gaussians centred around the mean RSSI $\mu_{s_id}$ of AP $d$ at state $s_i$ which is known only to the server.
By transforming into negative log-space and dropping the constants, we get $[p(r_t | s_i)] = \sum_{d=1}^{D}([r_{td}]-\mu_{s_id})^2$.
Since $[r_{td}]$ is encrypted, squaring requires one interactive multiplication.
To make this more efficient, the user also supplies $[r_{td}^2]$ upfront.
The server can then efficiently compute locally $[p(r_t|s_i)] = [r_{td}^2] -2\mu_{s_id}[r_{td}] + \mu_{s_id}^2$.

\textbf{Solving the optimzation problem.} 
We need to compute the costs $\vartheta_t(s_i)~\forall s_i \in S$ of placing a user at position $s_i$ in time step $t$ (Fig.\ \ref{fig:system}).
Computing the arguments to the min function requires only local additions and subtractions.
This leaves us with computing the $\min$ function (and the $\argmin$ to allow backtracking of the optimal path later).
While the basic problem of comparison has been widely studied in the literature, composing comparisons to efficient $\min$ and $\argmin$ algorithms has received less attention.
Proposed algorithms compute in a tree-like fashion pairwise comparisons, i.e., $[b] \leftarrow \text{\emph{Less-Than}}([x], [y])$, and then use the encrypted result $[b]$ in two interactive multiplications to select the smaller element as $[b] * [x] + (1-[b])*[y]$.
We achieve significant performance improvements by adapting the very efficient Multi-Party \emph{Less-Than} protocol proposed from \cite{kerschbaum2009performance} to the Two-Party case and additionally interleaving the comparisons with the minimum selection step.

\textbf{Updating the user's position:}
Finally, we run the $\argmin$ algorithm over all $[\vartheta_t(s_i)], s_i \in S$ to determine the most likely position of the user $[s^*_i]$ at time $t$ which is then, still under encryption, sent to the user.
The user, in possession of the private key, can then decrypt her position.

\section{Preliminary Results}
\label{sec:eval:accuracy}

We have prototypically implemented the proposed algorithm in Python 2.7 based on the GMP big integer library and the Paillier crypto system.
User and server communicate over a Gigabit LAN with a delay of $\approx 1 ms$.

\textbf{Processing and communication overheads.}
The overall complexity of the localization algorithm is $O(T \cdot N \cdot N')$ where $T$ is the number of time steps, $N$ is the number of possible states, and $N' $ is the number of predecessors per state.
Assuming realistic movement models, e.g., no transition through walls, no huge jumps, we get $N' << N$.
We have analyzed the average overheads of \emph{one} location update for different $N$ and $N'$ using a fixed number of $D = 20$ access points.
For $N = 160$ states with $N' = 5$ predecessors, one fresh position can be computed in roughly 10 seconds.
Increasing $N$ and $N'$, as expected, increases runtime approx.\ linearly.
Using more access points, i.e., increasing $D$, increases overheads only slightly since it impacts only local computations.
The communication overhead is in the order of a few MBs for the described setup.

\textbf{Localization Accuracy.} 
In principal, our approach can achieve the same accuracy as the original system \cite{viol2012hidden}, since the application of STPC does not per se introduce imprecision.
Practically, numeric precision and processing and communication overheads limit the accuracy.

Aliasgari et al.\ claim in \cite{aliasgari2013secure} that HMM computation requires precise floating point arithmetic to handle the involved products of small probabilities.
However, their proposed floating point primitives for two-/multi-party computation are computationally prohibitively expensive in the context of HMM computation.
Thus, instead of implementing real support for floating point arithmetic in STPC, we follow \cite{viol2012hidden} and apply a negative log-space transformation, which reduces the requirements for point arithmetic in our scenario.
As experiments show, accuracy does not significantly increase when considering more than five decimal places.
Such precision can be provided under STPC by simply scaling all inputs to the integers before encryption.

More limiting are the high computational costs of the STPC protocols, which bound the number of states we can handle without prohibitive delays.
However, the number of states directly determines accuracy.
The original system \cite{viol2012hidden} separates the space into $40\times40\times40~cm^3$ voxels amounting to thousands of states in their test setup, while we can only handle up to, e.g., 160 states within a reasonable delay of $10 s$.
Still, this number of states is enough to achieve room-level accuracy in medium sized indoor environments.

\textbf{Security and Privacy.} 
The used STPC protocols are secure against semi-honest adversaries.
User location privacy is preserved since the server never sees the user's inputs in clear.
Server privacy is protected since the user learns nothing about the HMM except a few mappings from observations to positions, which is inevitable in our scenario.

\section{Conclusion \& Outlook}
We have motivated and designed a privacy preserving state-of-the-art WiFi-based indoor localization system and, for the first time, presented an actual implementation of the Viterbi decoder using Secure Two-Party Computation. 
Our preliminary results indicate that our approach is feasible for achieving room-level accuracy with position updates in near real time.
The main issue is the performance of the algorithm.
We expect that using more efficient STPC primitives such as Garbled Circuits for the $\min$ and $\argmin$ algorithms can significantly increase performance and with it accuracy.

\small
\bibliographystyle{abbrv}

\end{document}